\documentclass[useAMS,usenatbib,usegraphicx,fleqn]{mn2e}
\usepackage{amsmath}

\begin{document}

\title[Analytic profile for prestellar cores]
{An analytic column density profile to fit prestellar cores}
\author[W. B. Dapp and S. Basu]{Wolf B. Dapp$^{1}$\thanks{E-mail:
wdapp@astro.uwo.ca (WD)} and Shantanu Basu$^{1}$,
\\
$^{1}$Department of Physics and Astronomy, The University of Western Ontario,
London, Ontario, N6A 3K7, Canada \\
}

\date{}

\maketitle

\label{firstpage}

\begin{abstract}

We present a new analytical three-parameter formula to fit observed column 
density profiles of prestellar cores. It represents a line-of-sight integral 
through a spherically symmetric or disc-like isothermal cloud. The underlying 
model resembles the Bonnor-Ebert model in that it features a flat central 
region leading into a power-law decline $\propto r^{-2}$ in density, and a 
well-defined outer radius. However, we do not assume that the cloud is in 
equilibrium, and can instead make qualitative statements about its dynamical 
state (expansion, equilibrium, collapse) using the size of the flat 
region as a proxy. Instead of having temperature as a fitting parameter, our 
model includes it as input, and thus avoids possible inconsistencies. It is 
significantly easier to fit to observational data than the Bonnor-Ebert sphere. 
We apply this model to L1689B and B68. We show that L1689B cannot be in 
equilibrium but instead appears to be collapsing, while our model verifies that 
B68 is not far from being a hydrostatic object.

\end{abstract}


\section{Introduction}

In recent years, observational advances have made it possible to measure
column density profiles in prestellar cores. Various methods are being used 
in the literature. \citet*{WT1999} measured mm continuum emission, 
while \citet*{Bacmann2000} utilized mid-IR absorption, 
and \citet*{Alves2001} measured dust extinction and reddening of the light 
of background stars in the near-IR. A fourth method is to use flux measurements 
in optically thin lines \citep*{Tafalla2002}. Most of the column density profiles 
measured in these ways show certain prominent common features: a central
flat region, followed by a power-law decline. Some cores additionally exhibit 
signatures of steepening of the profile, while some show a more-or-less smooth 
merger to some background value of the column density \citep{Bacmann2000}. 

One model often used to fit to such cores \citep*[e.g.,][]{Evans2001,Teixeira2005}
is the \textit{Bonnor-Ebert sphere} (hereafter BE sphere). This model assumes an 
isothermal sphere in equilibrium, acted upon only by gravity and thermal pressure, 
and bound by some external pressure \citep*{Bonnor1956,Ebert1955}. In the central 
region its density profile is flat with density $\approx \varrho _{\mathrm{c}} \equiv%
 \varrho \left( r=0 \right) $. The size of this region is proportional to the Jeans 
 length, $R_{\mathrm{J}} \propto c_{\mathrm{s}}/\sqrt{G\varrho_{\mathrm{c}} }$. 
It transitions into a power-law decline that approaches the
\textit{Singular Isothermal Sphere} (SIS) with $\varrho _{\mathrm{SIS}} %
=c_{\mathrm{s}}^{2} \left(2\pi Gr^{2}\right) ^{-1}$ for large radii \citep*{Shu1977}. 
The cloud is finally cut off at some finite radius, where external pressure forces 
match the internal forces. The BE model invokes the dimensionless radius parameter 
$\xi \equiv r \sqrt{4 \pi G \varrho _{\mathrm{c}}}/c_{\mathrm{s}}$. The value 
$\xi _{\mathrm{crit}}=6.45$ marks a dividing line such that clouds truncated 
at dimensionless radius $\xi _{\mathrm{s}}>\xi _{\mathrm{crit}}$ are in an unstable
equilibrium, and clouds with $\xi _{\mathrm{s}}<\xi _{\mathrm{crit}}$ are stable.

The column density profile of the BE sphere, obtained by integrating the volume 
density numerically along a line of sight, matches well with some observations 
\citep[B68;][]{Alves2001}. Depending on the parameters, the power-law decline 
$\propto r^{-1}$ in column density can be more or less pronounced, or even almost 
completely absent. The profile steepens at the edge because the line of sight 
through the sphere becomes shorter. This is a geometric effect which is present 
in all truncated models, unless the density \textit{increases} sufficiently 
fast with radius.

While physically motivated and reproducing features of several observed column 
density profiles, the BE model has shortcomings. The most important is the key 
assumption of equilibrium. In fact, most fits are found to be supercritical 
\citep{Teixeira2005}, representing unstable equilibria. These states are not 
expected to exist in reality, as any perturbation will send them to immediate 
collapse. On a more practical side, the procedure of fitting the BE model to 
observations is quite involved. The volume density is only available as a 
numerical solution, which then needs to be integrated (again numerically) to 
calculate the column density. Sometimes the fit demands temperatures \textit{well 
above} those measured for the centres of prestellar cores \citep*{Bacmann2000, AndreEtAl2003, KirkEtAl2005}, 
and predicted by detailed models of the thermodynamics within the core 
\citep*{Galli2002}. Finally, most of the observed cores are deeply embedded 
within their parent clouds, and the source of a bounding external pressure is 
not obvious. The often-cited example B68 \citep{Alves2001} is an exception, 
as it is thought to lie within a hot HII region. 

We stress that the generic features of the BE density profile are not unique to 
equilibrium situations \citep[see][]{KandoriEtAl2005}. The flat region with adjacent $r^{-2}$ density profile 
appears also in solutions of the hydrodynamic equations for gravitationally collapsing 
objects \citep{Larson1969}, as long as pressure is not completely 
negligible \citep*{ShuAdamsLizano1987}. The pressure gradient then
establishes a region where it nearly balances gravity. Here, the density 
is nearly constant on the scale of the local Jeans length, shrinking in size 
over time as the density increases. Inverted, this requires that $\varrho \propto%
r^{-2}$ in the outer profile that is left behind outside the central 
region \citep*[see][]{Basu1997}. 

There is no reason why the above two features (flat region with size of Jeans 
length and adjacent $r^{-2}$ density profile) should be present in a non-self-gravitating 
cloud. However, \citet*{BP2003} find that convergent turbulent flows, with and even 
without self-gravity (which are expected to have very different volume density profiles),
nevertheless yield column density profiles that resemble those of a BE sphere. There
are three reasons for this seemingly surprising result. The first one is the effect 
of smoothing the data, by angle-averaging, and also by integration along the line of 
sight. Secondly, the BE sphere leaves the modeller the freedom to fit the size of the 
central flat region by varying the temperature. Indeed, \citet{BP2003} fit their 
simulated cores with BE temperatures in the range of $5-60~\mathrm{K}$, despite their models 
being strictly isothermal at $T=11.3~\mathrm{K}$. Finally, the position of the outer 
radius cutoff is somewhat arbitrary. This often leaves large parts of profiles 
unfitted. In this paper, we argue that fitting a prestellar 
core profile at a \textit{set} temperature \textit{does} still allow one to distinguish 
between different models of internal structure.

We propose an analytic density profile reproducing the 
characteristics of not only isothermal equilibria \citep{Bonnor1956,Ebert1955}, 
but also non-equilibrium collapse solutions \citep*[e.g.,][]{Larson1969}, and many 
observed profiles \citep{Bacmann2000}. 
Within the margins of uncertainty it fits the observations as well as the BE 
model does. However, it possesses a closed-form expression for the column density, and 
is therefore very easy to fit. Furthermore, the temperature can be an input parameter 
instead of a fitting parameter, so that the model avoids some of the possible 
inconsistencies of fitting the BE model to either observations or simulation results. 
We can use our model to make some inference about the dynamical state of the core.

This paper is organized the following way: Section \ref{sec:SphericalGeometry} 
describes the spherical model and its parameters, while Section \ref{sec:DiskGeometry} 
presents a corresponding model for intrinsically flattened objects. In Section 
\ref{sec:Applications}, we apply our model to B68 and L1689B, 
and we summarize our results in Section \ref{sec:summary}. 

\section{Spherical Geometry}\label{sec:SphericalGeometry}

\subsection{Basic model}\label{subsec:model_spher}

The characteristics found in observed column density profiles and theoretically 
both in equilibrium and collapse solutions can be parametrized by a volume 
density more generic than the BE sphere. We propose using the profile%
\begin{equation}
		\varrho \left( r\right) =\left\{
		\begin{array}{cc}
				\varrho _{\mathrm{c}}a^{2}/\left( r^{2}+a^{2} \right) & r\leq R, \\
				0 & r>R,%
		\end{array}%
		\right.   
		\label{eq:Rho}
\end{equation}%
characterized by a central volume density $\varrho _{c}$ and truncated at some 
radius $R$. The parameter $a$ fits the size of the flat region in terms of the 
Jeans length, and is given by
\begin{equation}
		a=k\frac{c_{\mathrm{s}}}{\sqrt{G\varrho _{\mathrm{c}}}},  \label{eq:DefinitionA}
\end{equation}%
where $G$ is the gravitational constant and $k$ is a constant of proportionality. 
This profile is also mentioned in \citet*{King1962} and \citet{Tafalla2002}. 
The temperature $T$, which can be constrained observationally, 
is \textit{not} used as a fitting parameter. It enters through the value of the 
isothermal sound speed $c_{\mathrm{s}}=\sqrt{k _{\mathrm{B}}T/\mu }$. 
The Boltzmann constant is denoted by $k _{\mathrm{B}}$, and $\mu =2.33~m_{\mathrm{H}}$ 
is the mean mass of a particle, where $m_{\mathrm{H}}$ is the mass of a hydrogen atom. 
We assume a $10\%$ number fraction of helium.  

\begin{figure}
  \includegraphics[width=0.75\hsize]{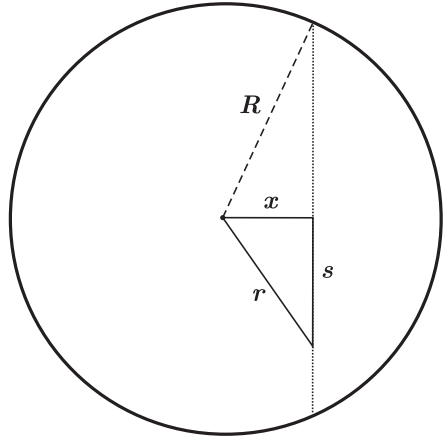}
      \centering
      \caption{
      	Schematic illustration of a cut through a spherical cloud of radius $R$. 
      	The observer is positioned along the direction of the coordinate $s$, and 
      	measures an integrated column density $\Sigma$ as a function of the offset 
      	$x$.
      }
      \label{fig:CloudCut}
\end{figure}

The column density is derived by integrating the volume density
along a line of sight through the sphere:
\begin{eqnarray}
		\Sigma \left( x\right) 
		&=&2\int_{0}^{\sqrt{R^{2}-x^{2}}} \varrho \left( s\right) ds \notag\\
		&=&2\int_{x}^{R}\frac{\varrho \left( r\right) rdr}{%
		\sqrt{r^{2}-x^{2}}} \label{eq:Integral_rho}
\end{eqnarray}%
where we have used the transformation $s=\sqrt{r^{2}-x^{2}}$ and hence $ds=rdr/\sqrt{r^{2}-x^{2}}$.
Fig. \ref{fig:CloudCut} defines the quantities appearing in this derivation.

Inserting equation (\ref{eq:Rho}) into equation (\ref{eq:Integral_rho}), we find that  
the integral is analytically tractable. The closed-form expression for the column 
density is then%
\begin{equation}
		\Sigma \left( x\right)=\frac{2 a ^{2}\varrho _{\mathrm{c}}}{\sqrt{x^{2}+a^{2}}}\arctan
		\left( \frac{\sqrt{R^{2}-x^{2}}}{\sqrt{x^{2}+a^{2}}}\right).
		\label{eq:Sigma_pre}
\end{equation}%
This can be re-written in terms of the central column density by introducing the 
ratio $c\equiv R/a$, called the concentration in \citet{King1962}. With 
$\Sigma \left( x=0\right)\equiv \Sigma _{\mathrm{c}} = 2a\varrho _{\mathrm{c}}%
\arctan \left( c\right)$ we find
\begin{eqnarray}
		\Sigma \left( x\right)
		&=&\frac{\Sigma _{\mathrm{c}}}{\sqrt{1+\left(x/a\right)^{2}}} \notag \\%
		&\quad& \times \left[\left.{{\arctan\left( \sqrt{\frac{c^{2}-\left(x/a\right)^{2}}{1+\left(x/a\right)^{2}}}%
		\right)} }\right/ {\arctan \left(c\right)}\right].
		\label{eq:Sigma}
\end{eqnarray}%

\begin{figure}
  \includegraphics[width=\hsize]{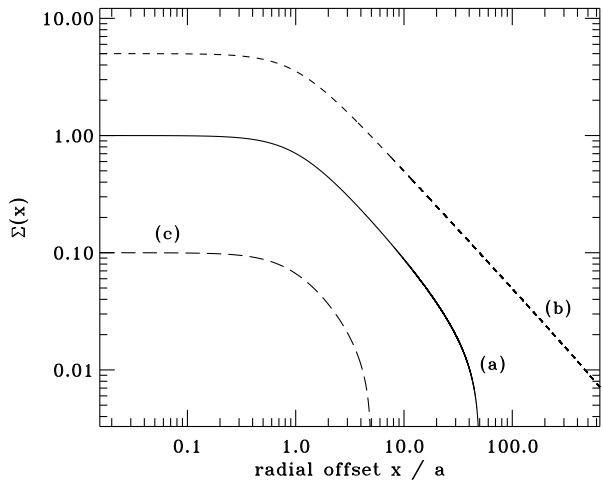}
      \caption{
				Plots of equation (\protect\ref{eq:Sigma}) for varying $c\equiv R/a$. 
				All cases have identical values for $a$. The central 
				column densities are chosen to be different in order for the plots not 
				to overlap. The solid line illustrates the three major characteristics 
				of this function: the flat region in the centre of size $a$, the adjacent 
				power-law decline $\propto x^{-1}$, a steepening at the edge for 
				geometric reasons (see text), out to the cut-off radius $R$. In this case, 
				$c=50$. The dashed-line shows a model dominated by the power-law regime, 
				with $c>1000$. $R$ is chosen so that the influence of the cut-off is
				negligible. The long-dashed line finally demonstrates that the power-law 
				can be suppressed entirely if $R \ga a$. In this case, $c=5$. 
      }
         \label{fig:SigmaGeneric}
\end{figure}

Fig. \ref{fig:SigmaGeneric} demonstrates that this profile possesses generic features 
that it shares with observations, collapse solutions, and with the BE model: a flat 
central region, a power-law decline, and steepening at the edge. We note that the effect 
of the boundary is exclusively contained in the factor in square brackets. 

The quantity $c$ determines the size of the region described by the power-law. If it is large, there is 
a pronounced power-law, whereas if $c$ approaches unity, the cut-off already dominates 
near the flat region and inhibits the power-law.

The three parameters to fit are (i) the outer radius $R$, (ii) the central
column density $\Sigma _{\mathrm{c}}$, and (iii) the size of the flat region 
$a$. The latter contains the product of $k$ and $\sqrt{T}$, as shown in equation 
(\ref{eq:DefinitionA}). If the temperature is pre-determined, $a$ only depends 
on the value of $k$. This then allows for a stability assessment, as demonstrated 
in the following section. 

\subsection{Model parameters}\label{subsec:params_spher}

The pressure in an isothermal system is given by $P=c_{\mathrm{s}}^{2}\varrho$,
and hence the pressure gradient for the spherically symmetric volume
density profile of equation (\ref{eq:Rho}) is%
\begin{equation}
		\frac{dP}{dr}=c_{\mathrm{s}}^{2}\frac{d\varrho }{dr}=\frac{-2c_{\mathrm{s}}
		^{2}\varrho _{\mathrm{c}}r}{a^{2}\left[ 1+\left( r/a\right) ^{2}\right] ^{2}}.
\end{equation}%
In the inner regions, where $r\ll a$,%
\begin{equation}
		\frac{dP}{dr}|_{r\ll a}=\frac{-2c_{\mathrm{s}}^{2}\varrho _{\mathrm{c}}r}{a^{2}}.
\end{equation}%
This demonstrates a simple point: the larger the flat region $a$, the smaller 
the pressure gradient in that region. The minimal value for $a$ is reached in an 
equilibrium situation. A larger $a$ would result in a smaller pressure gradient, 
gravity would win out, and collapse would ensue. Conversely, for $a$ smaller 
than its equilibrium value the pressure force would dominate, causing expansion. 
However, that case is not probable since a profile with $\varrho \propto r^{-2}$ 
mandates a strong gravitational influence. 

Another well-known model besides the SIS is the so-called Larson-Penston solution 
(LP solution), a self-similar spherical collapse solution, 
assuming a homogeneous initial density distribution \citep*{Larson1969,Penston1969}. 
This highly dynamical model does not assume equilibrium, and asymptotically reaches 
a density profile for which
\begin{equation}
		\varrho _{\mathrm{LP}}= 4.4~\varrho _{\mathrm{SIS}}
\end{equation}%
everywhere. We now want to find an expression for the parameter $a$ in the SIS and the
LP models. In the outer regions, equation (\ref{eq:Rho}) becomes%
\begin{equation}
		\varrho \left( r\gg a\right) \approx \frac{\varrho _{\mathrm{c}}a^{2}}{r^{2}} =%
 		\frac{\widetilde{k}c_{\mathrm{s}}^{2}}{2\pi Gr^{2}},
\end{equation}%
where $\widetilde{k}\equiv 2\pi k^{2}=1$ for the SIS (equilibrium) and $\widetilde{k}=4.4$ for the
asymptotic LP collapse solution. Therefore%
\begin{equation}
		k\equiv\sqrt{\frac{\widetilde{k}}{2\pi }}=\left\{
		\begin{array}{cc}
				0.399 & \text{for SIS}, \\
				0.837 & \text{for LP}.%
		\end{array}%
		\right.
\end{equation}%
This has a direct application to observed prestellar cores. If an estimate for 
the core temperature is available, equation (\ref{eq:Sigma}) can be fit to the 
column density profile, with $k$ as a fitting parameter. If $k \approx 0.4$, the 
system can be considered to be in equilibrium. If $k$ is significantly larger 
than $0.4$, and closer to $1$, the cloud under scrutiny is collapsing. 

\begin{figure}
  \includegraphics[width=\hsize]{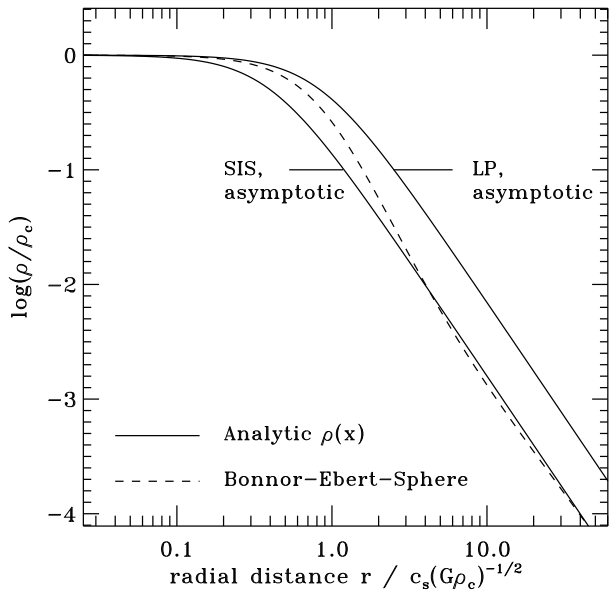}
      \caption{
				Comparison between the BE sphere (dashed line) and our model (solid lines) 
				for identical central temperature and cut-off radius. The top line is our 
				model with $k=0.837$, for which the profile approaches the asymptotic LP 
				solution. The lower line shows $k=0.399$, for which the SIS profile is reached. 
				The dashed line represents the BE model, which also asymptotically approaches 
				the SIS. 
      }
         \label{fig:BEvsSISvsLPvsGeneric}
\end{figure}

Fig. \ref{fig:BEvsSISvsLPvsGeneric} shows our model with $a$ chosen to converge 
to either the SIS or the asymptotic LP solution. We also plot the equilibrium BE 
model. In order to avoid boundary effects, the outer radius was moved 
to $\approx 10^{2}~c_{\mathrm{s}}/\sqrt{G\varrho_{\mathrm{c}}}$. The BE profile 
does not join onto the SIS profile right away. It overshoots, then steepens and 
asymptotically approaches the SIS profile at large radii. 

\begin{figure}
  \includegraphics[width=\hsize]{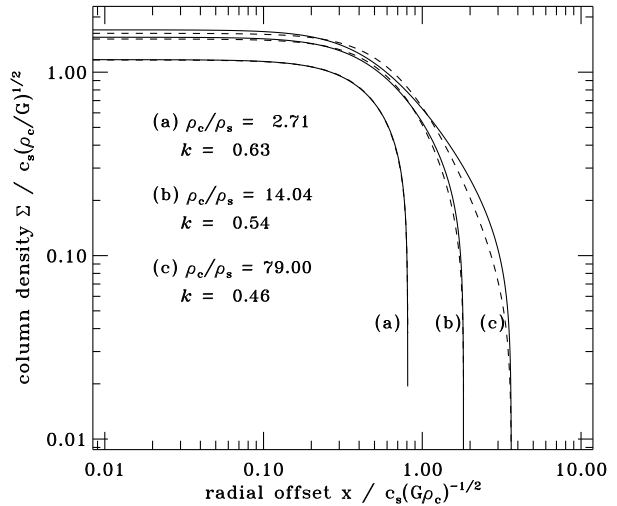}
      \caption{				
				Fit of the integrated column density of various BE spheres (dashed lines) 
				with our model (solid lines). 
				(a) is a subcritical (stable) model, while (b) shows a critical BE 
				sphere. In that case, $k=0.54$ (see text for explanation). (c) 
				presents a fit to a highly supercritical (unstable) BE model. 				
      }
      \label{fig:BEvsGeneric}
\end{figure}

We may ask which value of $k$ is most appropriate to model a critical BE sphere. 
The overshoot effectively produces a larger flat region if the solution is 
truncated at $\xi _{\mathrm{s}}=\xi _{\mathrm{crit}}$. The resultant $a$ has to 
be larger than for the SIS. Having fixed the temperature this can only be 
achieved by increasing $k$, as we show in Fig. \ref{fig:BEvsGeneric}. Profile 
(a) shows a subcritical BE sphere, i.e. a stable equilibrium solution. In this 
case, there is no discernible power-law region and the central flat region makes 
up a significant portion of the total radius. The cut-off becomes dominant just 
outside the flat region, and the best-fit dimensionless dynamics parameter is $k=0.63$.
Note that the density contrast is less than an \textit{e}-folding, which means 
a mean column density enhancement of $<40\%$ over the background. Such an object 
would not be observationally characterized as a prestellar core \citep*{AndreEtAl2009}. 

Profile (c) in Fig. \ref{fig:BEvsGeneric} shows the opposite situation, where the power-law portion of the profile is fit, 
and $k=0.46$ is closer to the equilibrium value. As shown in Fig. 
\ref{fig:BEvsSISvsLPvsGeneric}, the BE model initially has a power-law index 
steeper than $-2$. This is the reason why the profiles do not match up 
as well as for the other cases. Profile (b) finally shows a critical BE sphere. A 
density contrast greater than $\varrho _{\mathrm{c}}/ \varrho _{\mathrm{s}} = 14.0$ would 
characterize an unstable equilibrium, initiating collapse upon the smallest of 
perturbations. The two profiles are very similar, only minor deviations are 
discernible. In order to fit our model to the overshoot of the BE profile over the SIS, 
$a$ has to be larger, which can only be achieved by $k=0.54$. We demonstrate this effect 
on the examples of B68 and L1689B in Sections \ref{subsec:B68} and \ref{subsec:L1689B}: 
as a consistency check, we insert the derived BE temperatures into our model, and indeed 
find $k=0.57$ and $k=0.56$ respectively. Note that the BE model \textit{is} a 
valid model in absence of any effects other than gravity and thermal pressure, 
and thus we conclude that in this case, not the asymptotic value of $k=0.4$ but rather 
the critical BE value of $k=0.54$ is relevant for the stability assessment. We stress that if 
$k$ is significantly larger than that, it is strongly indicative of collapse.

Since the BE model effectively has $k \approx 0.5$ and collapsing clouds have 
$k \approx 1.0$, trying to force a BE model fit to such a cloud results in spurious 
higher temperatures. This can be seen in \citet{Bacmann2000}. A factor of $2$ in the 
dimensionless dynamics parameter $k$ needs to be compensated by a factor of $4$ in 
temperature when fitting a BE equilibrium state.

\subsection{Total mass}

Both our model and the BE sphere have a well-defined outer radius, and the total 
mass depends on it. For the profile of equation (\ref{eq:Rho}), the total mass has
a closed-form expression. Evaluating $\int_{0}^{R}4\pi%
r^{2}\varrho \left( r\right) dr$ yields%
\begin{equation}
		M=4\pi \varrho _{\mathrm{c}}a^{3}\left[ \frac{R}{a}-\arctan 
		\left( \frac{R}{a}\right) \right].   
		\label{eq:Mass}
\end{equation}%
There is evidence for some power-law behaviour in most observed profiles 
\citep[e.g.,][]{Bacmann2000}. In such a situation, the value of $R/a$ must be at 
least $\ga 10$ (cf. Fig. \ref{fig:SigmaGeneric}). Then, the contribution of $\arctan 
\left( c \right)$ is quite close to its limiting value of $\pi /2$, 
and the total mass is given by%
\begin{equation}
		M\approx 4\pi \varrho _{\mathrm{c}} a^{2}R = 4\pi k^{2}R c_{\mathrm{s}}^{2}/G
\end{equation}%
with $<15\%$ error. The equivalent expression for the BE model is 
(derived assuming an SIS profile at large radii)
\begin{equation}
		M _{\mathrm{BE}}\approx 2 R c_{\mathrm{s}}^{2}/G.
\end{equation}

Both expressions for the masses are linearly dependent on the unknown radius 
$R$. For a given (measured) temperature, our model additionally invokes the 
dimensionless dynamics parameter $k$, which is well constrained by the size 
of the flat region. The BE mass instead depends on the temperature of the fit, 
which we show in Sections \ref{subsec:params_spher} and \ref{subsec:L1689B} 
to potentially be a very poor value, and prone to overestimation by as much 
as a factor of $4$. 
 
\section{Disc geometry}\label{sec:DiskGeometry}

\subsection{Basic model}\label{subsec:model_disk}

Most observed cores do not appear circular in projection. A corresponding analysis 
as in Section \ref{subsec:params_spher} can be done for a disc-like geometry as well. 
Here, a generic face-on column density profile is%
\begin{equation}
		\sigma \left( r\right) =\left\{
		\begin{array}{cc}
				\sigma _{\mathrm{c}}\left/{\sqrt{1+\left(r/a\right) ^{2}}}\right. & r\leq R, \\
				0 & r>R.%
		\end{array}%
		\right.   
		\label{eq:disk_sigma}
\end{equation}%
This profile provides a good fit to the column density of collapsing 
flattened clouds \citep{Basu1997}, even though the models in that 
paper are not truncated.

Assuming vertical hydrostatic equilibrium and ignoring the effect of 
external pressure, the volume density is proportional to the square of 
the column density%
\begin{equation}
		c_{\mathrm{s}}^{2}\varrho =\frac{\pi }{2}G\sigma ^{2},
\end{equation}%
and the density accordingly is given by%
\begin{equation}
		\varrho \left( r\right) =\frac{\pi G}{2c_{\mathrm{s}}^{2}}\frac{\sigma _{%
		\mathrm{c}}^{2}}{1+\left( r/a\right) ^{2}}.
\end{equation}%
The assumption of vertical hydrostatic equilibrium is well justified by 
simulations when some source of flattening is present, such as rotation 
\citep*{Narita1984}, or magnetic fields \citep*{FiedlerMouschovias1993}. 
Abbreviating $\pi G \sigma%
_{\mathrm{c}}^{2}/\left(2c_{\mathrm{s}}^{2}\right) = \varrho _{\mathrm{c}}$, 
the volume density profile has exactly the same form as for the spherical case 
(with a different constant).

\subsection{Model parameters}\label{subsec:params_disk}

Integrating $\varrho \left( r\right)$ through the disc viewed edge-on, the 
column density versus the offset $x$ has the same functional form
as derived above for the spherical case in equation (\ref{eq:Sigma}). The only 
difference is that the central column density is now given by $\Sigma 
_{\mathrm{c}}=\pi k \sigma _{\mathrm{c}} \arctan \left( c\right)$, since 
for a thin disc, the relation between $a$ and the Jeans length is
\begin{equation}
		a=k\frac{c_{\mathrm{s}}^{2}}{G\sigma _{\mathrm{c}}}.
		\label{eq:Jeans_length_disk}
\end{equation}

This similarity means that both a flattened and a spherical object 
(see Section \ref{sec:SphericalGeometry}) can be fit with the same formula. 
Note that $\sigma \left( r\right)$ is the face-on column density, whereas 
$\Sigma \left( x\right)$ is its edge-on counterpart.

We can compare the (face-on) generic column density profile to one of
an equilibrium solution, in order to get the minimal value for $a$, as
done for the spherical case above. Here, the appropriate profile is the 
singular isothermal disc, characterized by%
\begin{equation}
		\sigma _{\mathrm{SID}}=\frac{c_{\mathrm{s}}^{2}}{2\pi Gr}.
\end{equation}%
Following the same steps as in Section \ref{subsec:params_spher}, the 
large-radius asymptote of equation (\ref{eq:disk_sigma}) is
\begin{equation}
		\sigma \left( r\gg a\right) \approx \frac{\sigma _{\mathrm{c}}a}{r},
\end{equation}%
yielding $a=c_{\mathrm{s}}^{2}/(2\pi G \sigma _{\mathrm{c}})$. Comparing 
this expression to equation (\ref{eq:Jeans_length_disk}), we find $k=1/2\pi=0.160$ 
for equilibrium.

\citet{SaigoHanawa1998} show that for an isothermal disc-like cloud 
during runaway collapse, the self-similar column density profile is 
given by $3.61~\sigma _{\mathrm{SID}}$, analogous to how the LP solution 
is over-dense compared to the SIS. This provides a good estimate for the value 
of the dimensionless dynamics parameter for a dynamically collapsing disc-like
cloud, $k = 0.57$. Flattened cloud cores best fit with values for the 
dimensionless dynamics parameter $k \approx 0.2$ are therefore close to 
equilibrium, while $k \approx 0.6$ is strongly indicative of dynamical collapse.

\subsubsection{Effect of magnetic fields}\label{subsubsec:magneticfields}

Inclusion of magnetic fields into a spherical model poses a problem: the 
magnetic field lines cannot be arranged in a spherically symmetric way. 
There has to be a preferred direction. However, in disc-like geometry, the 
axis perpendicular to the disc is preferred already, and can be  
chosen as the axis of orientation of a large-scale magnetic field. Starting 
with initially straight field lines through a collapsing astrophysical gas 
sphere, flattening ensues along this preferred direction. Matter 
can contract significantly along the field lines, but the Lorentz force will impede 
motions perpendicular to the field \citep{Mouschovias1976}.

In the limit of a magnetic field much stronger than the field of the ambient
cloud, \citet{Basu1997}, \citet{NakamuraHanawa1997} and \citet{ShuLi1997} 
showed that the contribution of the magnetic field can be 
folded into a force calculation of the collapse of a magnetized disc by 
magnetic pressure modifying the effective sound speed, and magnetic tension 
changing the effective gravitational constant. Then we can write
\citep[see][]{Basu1998}
\begin{equation}
		a=k\frac{c_{\mathrm{s}}^{2}\left( 1+2\mu ^{-2}\right) }{G\sigma _{%
		\mathrm{c}}\left( 1-\mu ^{-2}\right) }.
\end{equation}%
We use the standard definition of the mass-to-flux ratio in units of the critical 
value \citep{Nakano&Nakamura}:
\begin{equation}
		\mu=\frac{\sigma \left( r\right)}{B _{z} \left( r\right)} 2\pi \sqrt{G}.
\end{equation}%
Collapse requires $\mu >1$, otherwise the magnetic forces will dominate over 
gravity and stabilize the cloud, and evolution can only happen on long time 
scales by ambipolar diffusion \citep[see, e.g.,][]{ShuAdamsLizano1987}. 

We can absorb the modification into an effective $k$, writing it as 
$k _{\mathrm{eff}}=k \left(1+2\mu^{-2} \right)\left( 1-\mu^{-2}\right)^{-1}$.
The value $k=0.160$ as derived for the non-magnetic case remains the minimal 
value, while $k _{\mathrm{eff}}$ can exceed it by a factor of $1.1$ to $2$ for 
$\mu\approx 2-5$, which is a reasonable estimate for supercritical cores
\citep{CiolekMouschovias1994,BasuMouschovias1994}. This means 
$k _{\mathrm{eff}}\la 0.3$, which leaves it still a factor of $2$ 
smaller than the value for the dimensionless dynamics parameter for 
dynamical collapse ($k \approx 0.6$, as shown above), and thus clearly distinguishable. 

\subsubsection{Effect of rotation}\label{subsubsec:rotation}

Similar to the problem for magnetic fields, rotation cannot be considered in 
strict spherical symmetry. In fact, rotation can be the cause of disc-like 
geometry, as a collapsing rotating spherical cloud will flatten along the 
rotation axis before contracting in the radial direction. 

There exist self-similar solutions for the collapse of a rotating thin disc 
\citep{Narita1984, SaigoHanawa1998}. Following the discussion in \citet{Basu1997}, 
one can express the additional effect of rotation as an effective acceleration 
$a_{\mathrm{eff}}=a _{\mathrm{T}}\left(1+a _{\mathrm{C}}/a _{\mathrm{T}}\right)$, 
where $a _{\mathrm{T}}$ is the thermal acceleration, while $a _{\mathrm{C}}$ denotes 
the centrifugal acceleration. 
Assuming the column density profile of equation (\ref{eq:disk_sigma}) and 
proportionality between the specific angular momentum and enclosed mass, this leads to
\begin{equation}
		a _{\mathrm{C}}/a _{\mathrm{T}} \approx 3 \times 10^{-3}.
\label{eq:thermal_over_rotational_acc}
\end{equation}%
This number is computed for a background rotation rate of the molecular cloud 
$\Omega _{\mathrm{c}}=10^{-14}~\mathrm{rad~s^{-1}}$, a central column number density 
$N _{\mathrm{c}} = 10^{21}~\mathrm{cm}^{-2}$, and a temperature $T=10~\mathrm{K}$.  
Hence, we find that the effective radial acceleration opposing gravity exceeds the 
thermal acceleration by less than $1\%$. This shows that, unlike magnetic forces, 
rotation does not significantly modify the size of the flat region and hence the 
dimensionless dynamics parameter $k$.

\section{Applications}\label{sec:Applications}

We fit both the BE model and our model to observational data by determining the 
best-fitting parameters using a standard Levenberg-Marquardt least-squares minimization 
algorithm based on MINPACK.

\subsection{B68}\label{subsec:B68}

We fit our model to B68, the prime example in the literature 
for an extraordinarily good fit of the BE model to a (angular-averaged) column 
density profile measured in near-IR dust extinction \citep{Alves2001}. B68 is 
an isolated Bok globule, a small dark cloud which has been studied extensively. 
\citet{Alves2001} assume a distance of $125~\mathrm{pc}$ and quote a BE mass 
of $2.1~\mathrm{M}_{\odot }$, a temperature of $T=16~\mathrm{K}$, an outer 
radius of $12,500~\mathrm{au}$, and a dimensionless outer radius 
$\xi_{\mathrm{s}} =6.9\pm 0.2$. \citet*{HotzelEtAl2002} and \citet*{LaiEtAl2003} 
more recently updated some of these values by measuring the temperature to be 
$T=10\pm 1.2~\mathrm{K}$ and $T=11\pm 0.9~\mathrm{K}$, respectively, and estimated 
the cloud to be closer by about  $25~\mathrm{pc}$, placing it at the near side of 
the Ophiuchus complex. This reduces the outer radius and decreases the BE mass to 
$\approx 1~\mathrm{M}_{\odot }$, but does not change the value for $\xi_{\mathrm{s}}$, 
as that is determined by the shape of the profile.

\begin{figure}
  \includegraphics[width=\hsize]{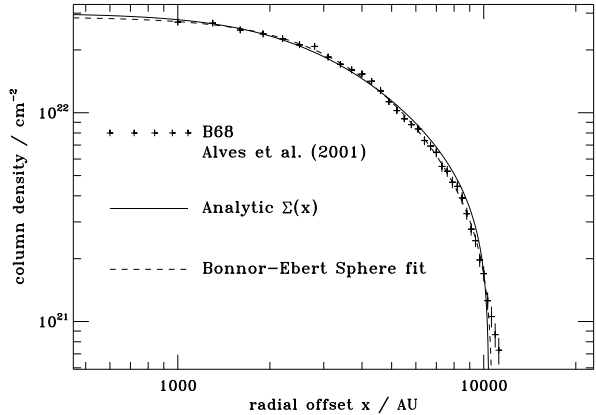}
      \caption{ 
				Best fit of the BE model (dashed line) and our model (solid line) 
				to the column density of B68 \protect\citep{Alves2001}. 
				Both fits follow the data points extremely closely and yield very similar 
				results for the total mass (see text). Like the BE model, our model fit
				is consistent with an equilibrium cloud.
      }
      \label{fig:FitB68}
\end{figure}

We perform a similar analysis as \citet{Alves2001}. We calculate a BE sphere, vary 
the temperature, the dimensionless and physical outer radii $\xi_{\mathrm{s}}$ and 
$R$, and fit the line-of-sight integral to the observational data, assuming a distance 
of $100~\mathrm{pc}$. This procedure yields a BE mass of $M=1.17~M_{\odot }$, 
a BE model temperature of $T=11.1~\mathrm{K}$, and a central number density of $n_{\mathrm{c}}=%
2.3\times 10^{5}~\mathrm{cm}^{-3}$. The outer radius of the best BE fit is 
$10,680~\mathrm{au}$, and the dimensionless outer radius $\xi_{\mathrm{s}} =7.0$. 
The best fit of our model to the same data yields a total mass of $M=1.2~M_{\odot }$, 
$R=10,420~\mathrm{au}$, and $n_{\mathrm{c}}=2.7\times 10^{5}%
~\mathrm{cm}^{-3}$. The size of the flat region is $a=2,830~\mathrm{au}$. Assuming 
a temperature of $T=11~\mathrm{K}$, as in the observations mentioned above, this
corresponds to $k=0.57$. We showed in Section 
\ref{subsec:params_spher} that a critical BE sphere is fit with $k=0.54$, and 
we conclude that the internal structure of B68 may indeed very closely resemble 
a critical BE sphere. Fig. \ref{fig:FitB68} shows that the best fits to 
B68 of the BE sphere and our model differ by very little over the whole range of data. 

The BE analysis is much more involved and computationally expensive than our 
model. It necessitates a numerical solution of the Lane-Emden equation. This 
ordinary differential equation (ODE) underlies the BE model, and does not have 
a general analytical solution. The numerical solution is truncated at the 
dimensionless radius $\xi_{\mathrm{s}}$ and converted into physical units. Only 
then can one integrate along lines of sight through the solution numerically, 
yielding the column density to be compared with observational data. 
In contrast, fitting our model to a dataset requires less than a dozen lines of 
code. 

\subsection{L1689B}\label{subsec:L1689B}

Another prestellar core to which we apply our model is L1689B. Its central column 
density is much greater and it is more extended than B68. Furthermore, 
it is not isolated, but rather embedded within the larger complex of the
Ophiuchus molecular cloud. Its profile was measured by \citet{Bacmann2000} 
using mid-IR absorption observations and
updated by \citet{AndreEtAl2003}. We use the East-West profile, which shows
evidence of steepening in the outer regions.

\begin{figure}
  \includegraphics[width=\hsize]{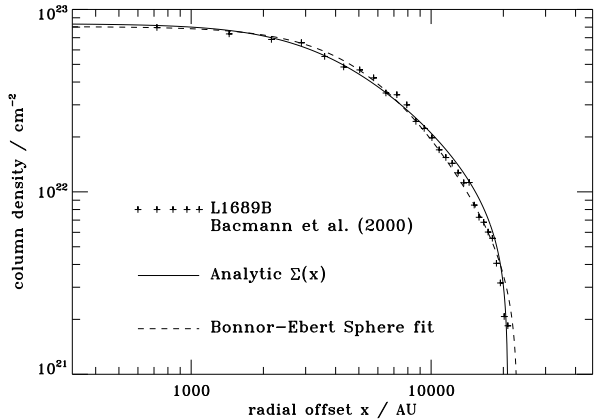}
      \caption{
				Best fit of the BE model (dashed line) and our model (solid line) 
				to the column density of L1689B, as measured by \protect\citet{AndreEtAl2003}. 
				Again, both fits are very similar and yield observationally almost 
				indistinguishable results. However, the BE model requires a relatively
				high temperature of $T=40~\mathrm{K}$, while our model can fit the data with 
				$T=10~\mathrm{K}$. The dimensionless dynamics parameter 
				$k=1.1$, which means that the cloud is collapsing. 
      }
         \label{fig:FitL1689B}
\end{figure}

The BE fit to the data in Fig. \ref{fig:FitL1689B} yields $\xi_{\mathrm{s}}%
=11.2$ (supercritical, unstable core) and a mass of $M=9.3~\mathrm{M}_{\odot }$, 
similar to what \citet{AndreEtAl2003} found. Fitting our model 
results in $M=9.1~\mathrm{M}_{\odot }$. However,
in contrast to the case of B68, the temperatures of the two fits are vastly 
different. The BE model requires a temperature $T=40~\mathrm{K}$, whereas 
our model can achieve the same quality of fit with a temperature $T=10~%
\mathrm{K}$. We fit the size of the flat region $a=3,600~\mathrm{au}$ which is a combination 
of the temperature and $k$. A temperature of $T=10~\mathrm{K}$ mandates 
$k=1.1$, which is very far from the value required for equilibrium, and 
even surpasses the value $k=0.837$ achieved in the highly dynamical asymptotic 
LP solution. As discussed in Section \ref{subsec:params_spher}, this is a 
strong indication that L1689B cannot be in equilibrium, but must instead 
be collapsing. 

Again, our model fits this dynamically evolving object just as well as it fit the
presumably almost static B68. It can do so without introducing inconsistent
temperatures, and with much less computational and coding effort.

\section{Summary and conclusions}\label{sec:summary}

We have introduced an analytic profile for the integrated line-of-sight 
column density of an isothermal spherical or flattened cloud. This cloud 
can either be in equilibrium or in a state of dynamical collapse. 
Our model is very simple to calculate compared to the BE model, 
and a few lines of code suffice to find a best-fit set of parameters. 
Another advantage lies in its ability to also encompass non-equilibrium states. 
The dimensionless dynamics parameter $k$ allows one to assess whether a cloud is 
near equilibrium or vigorously collapsing. At the same time, our model does not 
produce inconsistencies like the BE model regarding the object's temperature 
(which can be treated as a constrained quantity and not a free parameter). 

Our model can be applied both to spherical and to flattened clouds, and in both
cases yields the same functional form for the column density $\Sigma \left( x\right)$. 
It fits the size of the central flat region and allows the modeller to 
adjust either $k$ or the temperature to match it.

Results of fitting our model to B68 show that it is indeed a near-equilibrium 
cloud, with parameters very similar to the best-fit BE model. For L1689B, our 
model avoids the need for a high temperature (the BE model requires $T=40~\mathrm{K}$)
since it can be interpreted as having a temperature of $T=10~\mathrm{K}$ but 
being in a state of dynamical collapse. This finding is confirmed by the detection 
of infall motions for L1689B by means of the shape of the line profiles
in optically-thick molecular transitions \citep*{Bacmann2000,LeeEtAl2004}. In the future,
our model can be applied to many other prestellar cores. 

\section*{Acknowledgements}
We thank Joao Alves for giving access to his data on B68. We also thank Philippe 
Andr\'{e} for his data on L1689B and for his comments on the manuscript. SB was 
supported by a research grant from NSERC.

\end{document}